\documentclass[letterpaper,twocolumn,10pt]{article}
\usepackage{usenix2019_v3}

\usepackage{tikz}
\usepackage{amsmath}
\usepackage{subcaption}
\usepackage{xspace}

\usepackage{filecontents}

\usepackage{color, colortbl}
\definecolor{lightgray}{gray}{0.9}
\newcolumntype{C}[1]{>{\centering\let\newline\\\arraybackslash\hspace{0pt}}m{#1}}
\usepackage{tabularx,ragged2e,booktabs,caption}
\usepackage{footnote}
\makesavenoteenv{tabular}
\makesavenoteenv{table}

\usepackage{comment}

\newcommand{\supsym}[1]{\raisebox{6pt}{{\footnotesize #1}}}
\newfont{\coprimary}{phvr8t at 10pt}

\usepackage{pifont}
\usepackage{amssymb}
\newcommand{\cmark}{\ding{51}}%
\newcommand{\xmark}{\ding{55}}%

\newcommand{\name} {tinySDR\xspace}
\newcommand{\Name} {TinySDR\xspace}

\usepackage{enumitem}
\newcommand{\squishlist}{\begin{itemize}[itemsep=1pt,parsep=2pt,topsep=3pt,partopsep=0pt,leftmargin=0em, itemindent=1em,labelwidth=1em,labelsep=0.5em]}
\newcommand{\squishend}{\end{itemize}}
\usepackage[hang,flushmargin]{footmisc}

\usepackage{multirow}
\usepackage{multicol}

\newcommand{\xref}[1]{\S\ref{#1}}
\begin{document}
%-------------------------------------------------------------------------------

%don't want date printed
\date{}

\title{\LARGE \Name: Low-Power SDR Platform for \\ Over-the-Air Programmable IoT Testbeds\vspace{-25pt}}

%for single author (just remove % characters)
% \author{
% Paper \# 148, under submission NSDI 2020.
% % {\rm Your N.\ Here}\\
% % Your Institution
% % \and
% % {\rm Second Name}\\
% % Second Institution
% % copy the following lines to add more authors
% % \and
% % {\rm Name}\\
% %Name Institution
% } % end author

\author{%
Mehrdad Hessar\supsym{$\dagger$}, Ali Najafi\supsym{$\dagger$}, Vikram Iyer and Shyamnath Gollakota\\
{University of Washington}\\
\coprimary{\supsym{$\dagger$}Co-primary Student Authors}
}

\maketitle

{\bf Abstract ---} Wireless protocol design for IoT networks is an active area of research which has seen significant interest and developments in recent years. The research community is however handicapped by the lack of a flexible, easily deployable platform for prototyping {\it IoT endpoints} that would allow for ground up protocol development and investigation of how such protocols perform at scale. We introduce tinySDR, the first software-defined radio platform tailored to the needs of power-constrained IoT endpoints. TinySDR provides a standalone, fully programmable low power software-defined radio solution that can be duty cycled for battery operation like a real IoT endpoint, and more  importantly, can be programmed over the air to allow for large scale deployment. We present extensive evaluation of our platform showing it consumes as little as 30~uW of power in sleep mode, which is 10,000x lower than existing SDR platforms. We present two case studies by implementing LoRa  and  BLE beacons on the platform and  achieve sensitivities of -126~dBm and -94~dBm respectively while consuming 11\% and 3\% of the FPGA resources. Finally, using tinySDR, we explore the research question of whether an IoT device can demodulate concurrent LoRa transmissions in real-time, within its  power and computing  constraints.{\let\thefootnote\relax\footnote{{Accepted to  NSDI 2020.}}}
% \footnote{Accepted to  NSDI 2020}

%{\bf Abstract ---} Wireless protocols for IoT networks is an active area of research which has seen significant interest and developments in recent years. The research community is however handicapped by the lack of a flexible, easily deployable platform for prototyping {\it IoT endpoints} that would allow for ground up protocol development and investigation of how such protocols perform at scale. We introduce \name, the first software-defined radio platform tailored to the needs of power-constrained IoT endpoints. \Name provides a standalone, fully programmable low power software-defined radio solution that can be duty cycled for battery operation like a real IoT endpoint, and more  importantly, can be programmed over the air to allow for large scale deployment. We present extensive evaluation of our platform showing it consumes as little as 30~uW of power in sleep mode and can wirelessly program the FPGA and microcontroller, at a range of a university campus. We present two case studies by implementing LoRa  and  BLE beacons on the platform and  achieve sensitivities of -126~dBm and -94~dBm respectively while consuming 11\% or less of the FPGA resources. Finally, using \name, we affirmatively answer the research question of whether a IoT endpoint device can decode concurrent LoRa transmissions, within its  power, computation and delay  constraints.

\section{Introduction}
Recent years have seen development of numerous wireless protocols for Internet of Things (IoT) devices. In addition to longtime standards such as Bluetooth and Zigbee, a number of new protocols including LoRa, Sigfox, NB-IoT and LTE-M have been developed that achieve long ranges of more than a few  kilometers. Due to the lack of a de-facto standard, this space remains an active area of research for both industry and academia. The rapid advances in this space however present practical challenges for researchers: each of these protocols requires a dedicated radio chipset to evaluate, and these proprietary solutions often leave little room for protocol modification. The academic community is therefore {severely handicapped} by the lack of a \textit{flexible} platform, as even a complex multi-radio prototype cannot adapt to evaluate new protocols or even customize existing solutions. The current ecosystem therefore discourages researchers from investigating the important questions that arise when scaling up IoT networks, and more importantly taking a systematic approach to developing new protocols from the ground up.

% \begin{figure}[t]
%     \centering
%     \includegraphics[width=1\linewidth]{figs/map.eps}
%     \vskip -0.1in
%     \caption{\footnotesize{Our vision is to build an IoT testbed including hundreds of IoT endpoints spanning across a large area.}}
% 	\label{fig:vision}
%     \vskip -0.2in
% \end{figure}

%protocols for IoT remain an
%researchers are handicapped
%many protocols and active development
%recent interest
%new protocol every year
%802.11ah sigfox, nbiot lte-m, lora, ble zigbee
%each protocol is proprietary
%need a new testbed for each protocol
%more importantly  taking more systematic approach
%academic community is handicapped and can't develop own protocols

\begin{figure}[t]
\vskip -0.3in
    \centering
	\includegraphics[width=0.865\linewidth]{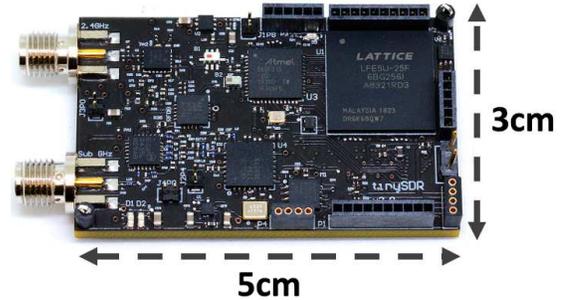}
	\vskip -0.1 in
	\caption{\footnotesize{{\bf \Name\ Hardware Platform.} It has two antenna ports for running IoT PHY and MAC protocols at 2.4~GHz and 900~MHz. {This image is the actual size of the platform on printed paper.}}}
	\label{fig:sdr_hardware}
	\vskip -0.2 in
\end{figure}

Ideally, we would like a large scale IoT network testbed with the flexibility to run \textit{any} IoT protocol at the PHY and MAC layers. Further,  since many of these IoT testbeds can span hundreds of endpoints across a  large campus or even a city, we need the ability to push changes to the PHY and MAC layers, using simple over-the-air software updates. This would allow for  performance comparisons on a single testbed to investigate the trade-offs between existing standards as well as showcase the advantages of an entirely new custom protocol. Moreover, to make such a system representative of real-world deployments, individual network nodes should model the constraints of IoT endpoints. Specifically, these devices should   have appropriate power controls and options to duty cycle transmissions, have an ultra-low power  sleep mode and also have interfaces to connect sensors. Finally, the ability to run these endpoints on batteries would also allow for flexibility of deployment in spaces without dedicated power access, or even in mobile scenarios.

\begin{table*}[t!]
    \centering
   % \vskip -0.1 in
    \footnotesize{
\begin{tabular}{|c|C{1.3cm}|c|c|c|C{1.1cm}|C{0.7cm}|c|C{1.1cm}|}
	\hline 
	\rowcolor{lightgray}
	{\bf Platform} & {\bf Sleep Power} & {\bf Standalone} & {\bf OTA} & {\bf Cost} & {\bf Max BW (MHz)} & {\bf ADC (bits)} & {\bf Frequency Spectrum (MHz)}& {\bf Size (cm)}\\
	{USRP E310 \cite{usrpe310, ad9361}} & 2820~mW & \cmark & \xmark & \$3000 & 30.72 & 12 & 70$\sim$6000 & 6.8$\times$13.3\\
	\hline
	{USRP B200mini \cite{usrpb200mini, ad9364}} & N/A & \xmark & \xmark & \$733 & 30.72 & 12 & 70$\sim$6000 & 5$\times$8.3\\
	\hline
    {bladeRF 2.0 \cite{bladerf2, ad9361}} & 717~mW & \cmark & \xmark & \$720 & 30.72 & 12 & 47$\sim$6000 & 6.3$\times$12.7\\
	\hline
	{LimeSDR Mini \cite{limesdr, limesdrmini, lms7002m}} & N/A & \xmark & \xmark & \$159 & 30.72 & 12 & 10$\sim$3500 & 3.1$\times$6.9\\
	\hline
	{PlutoSDR \cite{ad9363}} & N/A & \xmark & \xmark & \$149 & 20 & 12 & 325$\sim$3800 & 7.9$\times$11.7\\
	\hline
	{$\mu$SDR \cite{max2831, ad9228, max5189}} & 320~mW & \cmark & \xmark & \$150 & 40 & 8 & 2400$\sim$2500 & 7$\times$14.5\\
	\hline
	{GalioT \cite{galiot, rtl2832}} & 350~mW & \cmark & \xmark & \$60 & 14.4 & 8 & 0.5$\sim$1766 & 2.5$\times$7\\
	\hline
	{\bf{\Name}} & {\bf 0.03~mW} & {\bf \cmark} & {\bf \cmark} & {\bf \$55} & {\bf 4} & {\bf 13} & {\bf 389.5$\sim$510, 779$\sim$1020, 2400$\sim$2483} & {\bf 3$\times$5}\\
	\hline
\end{tabular}
    \vskip -0.1in
    \caption{\footnotesize{\bf Comparison Between Different SDR Platforms.} Costs are based on sale prices for commercial products without a public bill of materials (BOM) and published BOM prices for research prototypes. OTA refers to over-the-air programming capabilities.}
    \label{tab:platforms}
    \vskip -0.2in
}
\end{table*}

Realizing this vision however is challenging with existing software defined radio (SDR) platforms. Specifically, we require an SDR for  the flexibility of implementing different PHY protocols; but there is currently no SDR platform that meets the requirements of IoT endpoints (see Table~\ref{tab:platforms}). Existing SDR systems  consume  large amounts of power for transmitting data, do not support ultra-low power  sleep modes, require wired infrastructure and often a dedicated computer and furthermore, are expensive. More importantly, none of the existing SDR platforms support over-the-air programming to update PHY or MAC protocols.  Finally, IoT devices prioritize power consumption and communication range and hence use limited radio bandwidth --- LoRa, Sigfox, NB-IoT, LTE-M, Bluetooth and ZigBee use only 500~kHz, 200~Hz, 180~kHz, 1.4~MHz, 2~MHz and 2~MHz respectively. In contrast,  existing SDR  platforms focus on achieving high performance in terms of bandwidth because {\it they are tailored to the needs of gateway devices and not for IoT endpoint devices.}

%ideally large network testbed
%wirelessly update, small modifications
%test performance, etc
%custom protocols
%within constraints of endpoint

%need development platform
%while there has been work on access points SDRs/gateways, there is no sdr for Iot endpoints
%everything that exists is high power, most require wired infrastructure
%need wired backbone to update
%cost

Driven by a need for such a platform in our own research, we design \name as shown in Fig.~\ref{fig:sdr_hardware}, the first SDR platform tailored to the needs of IoT endpoints. TinySDR provides an entirely standalone solution that incorporates a radio front-end, FPGA and microcontroller for custom processing, over-the-air FPGA and microcontroller programming capabilities, a microSD card interface for storage, ultra-low power sleep modes and highly granular power management options to enable battery-powered operation. It is capable of transmitting and receiving in both the 900~MHz and 2.4~GHz ISM bands, supports 4~MHz of bandwidth which is  sufficient for most IoT protocols including Bluetooth, Zigbee, LoRa, Sigfox, NB-IoT and LTE-M, and can achieve the high sensitivities of commercial solutions such as LoRa chip~\cite{sx1276}. Additionally it includes multiple analog and digital I/O options for connecting sensors. % and measures only $3\times5~cm$.

%To do this, we systematically analyze the bandwidth and sensitivity requirements for IoT endpoints and design a real-time SDR that meets the needs of existing  IoT protocols  while ensuring flexibility at the PHY and MAC layers (see~\xref{sec:hardware}).

Designing such a SDR platform required addressing multiple  systems, architecture, power and engineering challenges:  
\squishlist
    \item {\bf Low-power hardware architecture.}  
    Achieving a small form-factor, low-power SDR requires a minimalist design approach that can satisfy the real-time needs of IoT protocols and ensure flexibility at the PHY and MAC layers. To do this, we exploit recent advances in small, low-power microcontrollers, FPGAs and flash memory  to pick the right components for our platform (see~\xref{sec:hardware}). We use a low-power  FPGA to run the PHY layer while the microcontroller runs the MAC protocols  as well as handles the I/O operations between the FPGA, radio, memory and sensor interfaces (see~\xref{sec:interface}).
    \item {\bf Efficient power management.}  Achieving highly granular power management needed for battery-powered operation and enabling ultra-low power sleep modes requires shutting down parts of SDR when not in use. This is important for IoT endpoints that {perform} duty-cycle operations and require an ultra-low power sleep mode to achieve a long battery life. This presents a design tradeoff between the complexity of toggling the power of each hardware component ON and OFF, and the cost of additional circuitry to do so. We address this challenge in~\xref{sec:powermanagement} and achieve  sleep power as low as 30~$\mu$W.
     \item {\bf Over-the-air SDR programming.} Enabling a truly scalable system requires the ability to update the PHY and MAC layers on the platform, over-the-air, in a testbed deployment.  This however also introduces the challenge of over-the-air FPGA and microcontroller programming as well as communicating these updates robustly to each device in the network while minimizing power consumption and network utilization. We use a dedicated wireless backbone subsystem complete with a MAC protocol and its own flash memory to program both the microcontroller and FPGA. Additionally we leverage compression and low-power decompression algorithms to minimize  network  downtime during the updates (see~\xref{sec:ota}) 
     
\squishend

Fig.~\ref{fig:platform_power} shows the power consumption of the radio module in \name\ compared to existing SDR platforms. We evaluate \name's performance  by presenting case studies of two common protocols: LoRa and BLE beacons, and also evaluate \name\ in a campus-testbed of 20 devices. 
\squishlist
\item LoRa modulation and demodulation use  4\% and 11\% of the FPGA resources respectively and achieve a sensitivity of -126~dBm for 3.12~kbps, which is similar to an SX1276~\cite{sx1276} LoRa chip with the same configuration. Further, the FPGA supports real-time modulation and demodulation of all LoRa spreading factors from 6 to 12. A LoRa MAC implementation on our MCU is compatible with the {\it The Things Network}.

\item \Name supports 2.4 GHz BLE beacon transmissions. The full baseband packet generation on the FPGA uses  3\% of its resources. The platform can perform frequency hopping with a delay of 220~us and  achieves a sensitivity of -94~dBm which is comparable to the commercial BLE chipsets~\cite{cc2650}.
\squishend

%\textcolor{red}{In addition to characterizing the performance of our system we explore how the unique capabilities of \name could be used to answer new research questions. Researchers have shown interest in performing concurrent transmissions for protocols such as LoRa~\cite{lorasigcomm17, netscatter} to improve network capacity. Because LoRa transmitters with different bandwidths will generate chirps with different slopes, we can exploit this to decode concurrent packets. We demonstrate that \name is capable of decoding two simultaneous LoRa transmissions with different bandwidths using a custom decoder implemented on our FPGA.}

{Finally, we present a case study of how the unique capabilities of \name could be used to answer new research questions. Recent work has explored techniques to enable concurrent transmissions in LoRa networks~\cite{lorasigcomm17, netscatter}; however these solutions were prototyped on USRPs and it is unclear if IoT endpoints can decode concurrent transmissions in real-time within their power and resource constraints. We implement a custom decoder on \name to demonstrate for the first time that IoT endpoints \textit{can} receive concurrent transmissions.}

{\bf Contributions. }To summarize, we design the first SDR platform tailored to the needs of IoT endpoint devices. By making careful design and architectural choices, our platform achieves low power, supports IoT protocols at both 900~MHz and 2.4~GHz and has  computation resources to do on-board processing. We present a highly granular power management scheme that enables duty-cycled operation and 10,000x lower power sleep modes. {We also develop the first over-the-air SDR programming capability to support PHY and MAC updates in a wireless testbed.} We characterize and evaluate our platform with case studies of LoRa and BLE beacons.  Finally, we present a research exploration of concurrently receiving multiple LoRa transmissions on our SDR platform.

{\bf Platform availability.} We will release the \name platform, along with its hardware schematics and software, for others to use and contribute to, before the conference.

\section{SDR Requirements for IoT Nodes} 
\begin{figure}[t]
    \centering
    \includegraphics[width=1.03\linewidth]{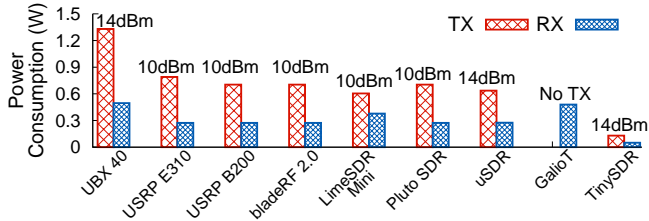}
    \vskip -0.1in
    \caption{\footnotesize{{\bf Radio Module Power Consumption for Each Platform.} The TX output power of each radio module is shown on top of it.}}
	\label{fig:platform_power}
    \vskip -0.2in
\end{figure}

To motivate the need for \name and inform our design decisions, we begin by identifying the key requirements for an IoT endpoint. These include 1) operation in the 900 MHz and 2.4~GHz bands, 2) low power operation which requires the ability to transition to ultra-low power sleep mode, 3) standalone operation which requires an on-board control unit to duty cycle the radio, 4) over-the-air programming capabilities for large scale IoT testbeds, 5) low cost per node, and 6) at least 2~MHz bandwidth to support IoT  protocols including LoRa, SIGFOX, LTE-M, NB-IoT, ZigBee and Bluetooth. While there are a number of commercially available SDRs such as the USRP, BladeRF, PlutoSDR, and LimeSDR~\cite{usrpe310,bladerf2,usrpX300,limesdrmini,plutoSDR} on the market and  SDR research prototypes such as WARP, Argos, SORA, SODA, KUAR, Tick, $\mu$SDR, OpenMili, and GalioT~\cite{amiri2007warp,khattab2008warp,anand2010warplab,tan2011sora,guddeti2019sweepsense,galiot,musdr,kuo2012compact,dutta2010putting,zhang2016openmili,soda,wu2017tick, minden2007kuar,sutton2010iris,ng2010airblue,argo1,argo2}, all of them are designed as {\it gateway devices} and do not satisfy many of the above constraints. Here, we analyze the shortcomings of  these platforms in the context of these requirements.
\squishlist
\item {\bf Low power operation and sleep mode.} Fig.~\ref{fig:platform_power} compares the power consumption of the radio module {\it alone} in existing SDR platforms, since each one has different peripherals. We find that most SDR platforms consume 200-300~mW in receive mode, but a lot  more power when transmitting. While this may be acceptable for a gateway devices that are more often receiving, typical IoT endpoints do the opposite and are required to transmit data like sensor information. Moreover, real IoT nodes spend a very short time transmitting before transitioning to ultra-low power sleep modes. Although IoT radios often consume tens to hundreds of milliwatts of power, the key to achieving long battery lifetimes is exploiting their microwatt power sleep modes. Table~\ref{tab:platforms} shows that none of the other platforms can benefit from duty cycling as they consume more power in sleep mode than \name does when transmitting; \name's microwatt power consumption in sleep mode enables dramatic power savings with duty cycling.

\item{\bf Standalone operation and cost.} We  observe that some of these platforms do not allow for standalone operation, i.e., they cannot be used in a testbed deployment without an external computer. Among the ones that do, the Embedded USRP and bladeRF cost \$700 or more per unit making large scale deployments expensive. $\mu$SDR allows for standalone operation but only operates at 2.4~GHz and cannot support protocols like LoRa. GalioT~\cite{galiot} uses the low cost RTL2832U radio~\cite{rtl2832} connected to a Raspberry Pi computer which allows for standalone operation, however it does not support 2.4~GHz band. Moreover, this platform is \textit{receiver only} and cannot be used to prototype a typical IoT node that transmits data. %This leaves \name as the only low cost platform that can transmit and receive in both ISM bands without being tethered to an external computer.

\item{\bf Over-the-air (OTA) programming.} As shown in Table~\ref{tab:platforms}, all existing SDR platforms rely on wired interfaces for programming. This means that even if one of these systems were connected to a battery, running an experiment would require either tethering each one to a wired network or individually programming them. An OTA programming system is crucial to realizing the goal of a large scale wide area testbed as without it, researchers have to decide between limiting themselves to deployment scenarios with wired infrastructure that are not representative of real IoT use cases or traveling over \textit{kilometer} distances to update individual nodes for \textit{each} minor protocol modification, which would be unmanageable at  scale.
\squishend
\vspace{-0.2 in}
\section{\Name\ Platform}
\begin{figure*}[t]
    \centering
	\includegraphics[width=0.8\linewidth]{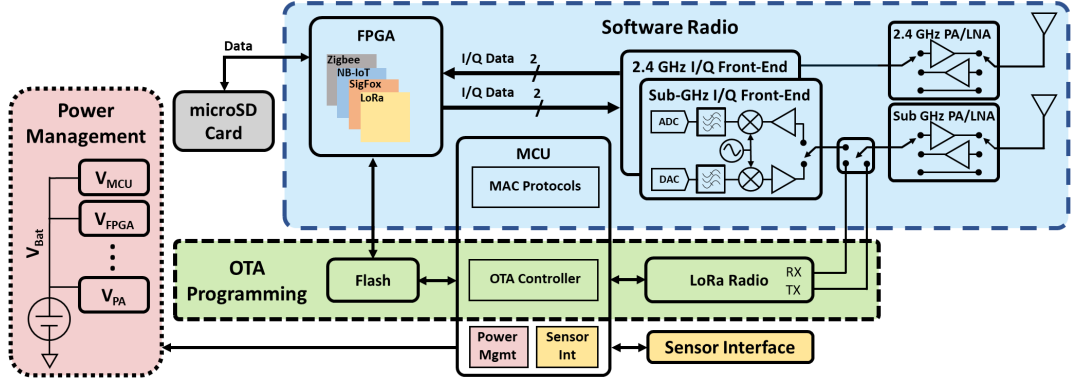}
	\vskip -0.15in
	\caption{\footnotesize{{\bf \Name System Block Diagram.} A complete system diagram showing all of the components of \name. This includes the software radio consisting of the radio, amplifiers, and FPGA, OTA programmer which uses a LoRa radio and flash memory to store programs, and a power managment system with the flexibility to turn off power consuming components. Each of these subsystems are controlled in software running on the MCU.}}
	\vskip -0.23in
	\label{fig:system}
\end{figure*}

We first describe our design choices for the different components of our hardware shown in Fig.~\ref{fig:system} and explain the interfaces between them. Next we present the power management module which enables our ultra-low-power sleep mode. Finally, we describe our over-the-air update protocol including  decompression algorithms and over-the-air reprogramming. 

\vspace{-0.2 in}
\subsection{Hardware Design}
\label{sec:hardware}
We seek to minimize power consumption and cost while offering the flexibility of an SDR to process raw samples. %Here, we outline the system architecture along with a detailed analysis of our design decisions.

\subsubsection{Designing the Software Radio}
The core block on our platform is the software-defined radio, a programmable PHY layer that processes and converts bits to radio signals and vice versa. We begin by explaining our choices for the primary components of an SDR which are a radio chip that provides an interface for sending and receiving raw samples of an RF signal as well as an FPGA that can process these signals in real time. We then discuss the supporting peripherals for these devices such as a power amplifier (PA) to boost the output of the radio chip and non-volatile memory for the FPGA to read and write data from.

{\bf Choosing a radio chip.}   We begin by choosing a radio chip as its specs define the requirements for the FPGA and other blocks. Our primary requirement is that the chip supports reading and writing raw complex I/Q samples of the RF signal. As shown in Table~\ref{tab:radios}, current SDR systems use I/Q radio chips that are designed to cover a multi-GHz spectrum and have high ADC/DAC sampling rates to support large bandwidth. For example, AD936x~\cite{ad9361} series which is used in USRP and ADPluto can transmit from 325~MHz to 3.8~GHz and supports sampling rates as high as tens of MHz. Each of these specs such as wide bandwidth, low noise, and high sampling rate represent fundamental trade offs of power for performance, and therefore these chips consume watts of power. Moreover, some of these radio chips costs more than \$100. 

We instead take a different approach: identify the minimum required specs and find a radio that supports them. Specifically, an IoT platform must be able to operate in at least the 900~MHz and 2.4~GHz ISM bands, have 4~MHz of bandwidth, while otherwise minimizing power and ideally costing less than \$10. We analyze all of the commercially available radio chips that provide baseband I/Q samples and list them in Table~\ref{tab:radios}, where only the AT86RF215 supports all of our requirements. In addition to lower cost and support for both frequency bands, it also consumes less power than the MAX2831 and the SX1257. Moreover, the AT86RF215 integrates all the necessary blocks including an LNA, programmable receive gain, {automatic gain control (AGC)} and low pass filter, ADC on the RX chain, as well as a DAC and programmable PA with a maximum power of 14~dBm on the TX side.  In terms of noise, the RF front-end has a 3-5~dB noise figure which is even better than the noise figure of the front-end used in Semtech SX1276 LoRa chipset, suggesting it should be able to achieve long range performance. It consumes 5x less power than the radios used on other SDRs as shown in Fig.~\ref{fig:platform_power} and has built in support for common modulations such as MR-FSK, MR-OFDM, MR-O-QPSK and O-QPSK that can save FPGA resources or power by bypassing the FPGA entirely.

%The receiver chain passes signal through an LNA with noise figure of 4.5 and then downconverts the signal using a mixer that is controlled by frequency synthesizer (PLL).

\begin{table}[t]
    \centering
   % \vskip -0.1in
    \caption{\footnotesize{\bf Existing Off-the-Shelf I/Q Radio Modules.}}
    \vskip -0.15 in
    \footnotesize{
    \begin{tabular}{|C{2cm}|C{2.2cm}|c|c|}
         \hline
         \cellcolor{lightgray}{\bf I/Q Radio} & \cellcolor{lightgray}{\bf Frequency (MHz)} & \cellcolor{lightgray}{\bf RX Power (mW)} & \cellcolor{lightgray}{\bf Cost}\\
         \hline
         AD9361~\cite{ad9361} & 70$\sim$6000 & 262 & \$282\\
         \hline
         AD9363~\cite{ad9363} & 325$\sim$3800 & 262 & \$123\\
         \hline
         AD9364~\cite{ad9364} & 70$\sim$6000 & 262 & \$210\\
         \hline
         LMS7002M~\cite{lms7002m} & 10$\sim$3500 & 378 & \$110\\ 
         \hline
         MAX2831~\cite{max2831} & 2400$\sim$2500 & 276 & \$9\\ 
         \hline
         SX1257~\cite{sx1257} & 862$\sim$1020 & 54 & \$7.5\\ 
         \hline
         {\bf AT86RF215~\cite{at86rf215}} & {\bf 389.5$\sim$510 \newline 779$\sim$1020 \newline 2400$\sim$2483} & {\bf 50} & {\bf \$5.5}\\ 
         \hline
    \end{tabular}
    }
    \vskip -0.3in
    \label{tab:radios}
\end{table}

{\bf Picking an FPGA.} Now that we have chosen a radio chip, the next step in our design process is to find an FPGA that can interface with it. Aside from minimizing power {and cost}, we would also like to maintain a small form factor {and short wake-up time}. {Although flash-based FPGAs are capable of fast wake-ups, they are more expensive compared to SRAM-based FPGAs with the same number of logic elements.} We use LFE5U-25F~\cite{latticeLFE5U} FPGA from Lattice Semiconductor for baseband processing which is an SRAM-based and has 24~k logic units. This chip provides a greater number of look up tables (LUTs) than the FPGAs on the PlutoSDR and LimeSDR mini, and at lower cost.

% and so we also consider the physical package size of the FPGA.

%We also choose a small package of this FPGA to decrease the platform's size.

% Maybe add a comparison of the Microsemi FPGA you used on V1?

{\bf Adding a power amplifier (PA).} AT86RF215 only supports a maximum transmit power of 14~dBm which is traditionally used by IoT radios but is less than the 30~dBm maximum allowed by the FCC. To provide flexibility, we add optional PAs. Given the high cost and power requirements of wide-band PAs that could operate at both 900~MHz and 2.4~GHz we instead select two different chips: the SE2435L~\cite{SE2435L} for 900~MHz and SKY66112~\cite{SKY66112} for 2.4~GHz. Our 900~MHz PA supports up to 30~dBm output power, and the 2.4~GHz PA can output up to 27~dBm. Both chips also include an LNA for receive mode and a built in circuit to bypass either of these components for power savings. In receive mode, we can either pass the incoming signal through the LNA and then connect it to the radio or completely bypass the LNA and connect the signal directly. The maximum bypass current is 280 uA and the sleep current of both power amplifiers is only 1 uA. In transmit operation we can pass the signal through the PA and amplify the signal or turn off the PA and pass the signal directly to the antenna for transmit power $<$ 14~dBm.

{\bf Picking the microcontroller.} We use a microcontroller to control all the individual chips and toggle all of these power saving options. In addition to having a low sleep current it must be able to support multiple control interfaces, have enough memory resources to support IoT MAC protocols and also be able to run a decompression algorithm for our OTA system. We select the MSP432P401R~\cite{msp432P401R} a 32-Bit Cortex M4F MCU which meets all of our requirements with  less than 1~uA sleep current, has 64~KB of onboard SRAM and 256~KB of onboard flash memory.  In addition to controlling the I/Q and backbone radio parameters, and reprogramming of the FPGA, the MCU performs the important function of power management. It is responsible for toggling ON and OFF the power amplifiers, as well as performing power-gating by turning ON and OFF different voltage regulators in~\xref{sec:powermanagement}.

\subsubsection{Designing OTA Update Hardware}
While the above discussion enables a small, low power, low cost SDR for easy deployment, FPGAs and microcontrollers typically require a wired interface for reprogramming. Here we present the hardware for the OTA update system to reconfigure and program \name nodes wirelessly. 

{\bf OTA wireless chipset.} A key question when designing an OTA update system is, what wireless protocol should be used? To support wide area networking, we focus on protocols designed for long range operation. We analyze all of the available long range protocols and select LoRa for our OTA system for a number of reasons. First, LoRa receivers have a high sensitivity which enables kilometer ranges. LoRa also support a wide range of data rates from 11~bps to 37~kbps which allows us to trade off rate for range depending on the deployment scenario. Moreover, LoRa is becoming more and more wide-spread in the US. We use the SX1276 Semtech chipset~\cite{sx1276} which is available for \$4.5, minimizing cost.

% For example with 30~dBm output power, $SF = xxx$, $BW = xxx$ and a 2~dBi antenna, we can achieve xxx~kbps at xxxx~km. 
% \noindent{\bf OTA MAC.} In \xref{sec:OTA}, we show that the update takes about 1 minute to finish depending on the update rate and the update bitstream file size. Moreover, we provision only one whole network firmware update per day. So, to update all the devices' firmware, each device goes to update mode at a certain time during the day. \textcolor{red}{explain the update MAC more}XXXX

{\bf Flash Memory.} Our FPGA  is SRAM based and does not include on-chip non-volatile memory for storing programming data. We instead store the firmware bitstream on a separate flash memory chip. The FPGA programming bitstream is 579~KB and the MCU programs require a maximum of 256~KB. We chose the MX25R6435F flash chip with 8~MB memory. Although this is far more than the size required, it allows \name to store multiple FPGA bitstreams and MCU programs to quickly switch between stored protocols  without having to re-send the programming data over the air.

\vspace{-0.05 in}
\subsection{Interfacing Between Blocks}
\label{sec:interface}
%%%In this section, we explain the operation of each of the above blocks in detail. To do this we walk through the technical challenges of reading the digitized signal into the FPGA at the full sampling rate of the radio output and into internal memory followed by saving them to an SD card. Next we give an end to end description of our over-the-air update system including the access point communication protocol, compression and reprogramming sequence.

%In order to receive samples he interfaces between each of the components described above as well as their internal software blocks. We 

%We first explain our I/Q interface with I/Q radio. Then we explain our memory interface and finally we show steps of our over-the-air system to update each \name node in the network.

\subsubsection{Reading and Writing I/Q Samples}
The AT86RF215 radio chipset samples baseband signals at 4~MHz with a 13~bit resolution for both I and Q. Operating at the full rate therefore requires an interface which can support a throughput of over 100~Mbps without consuming a large amount of power to meet our design objectives. To do this we use  low-voltage differential signaling (LVDS)~\cite{lvdsTechnology} which is a high-speed digital interface that reduces power by using lower voltage signals but maintains good SNR by sending data over two differential lines to reduce common mode noise.

\begin{figure}[t]
    \centering
	\includegraphics[width=\linewidth]{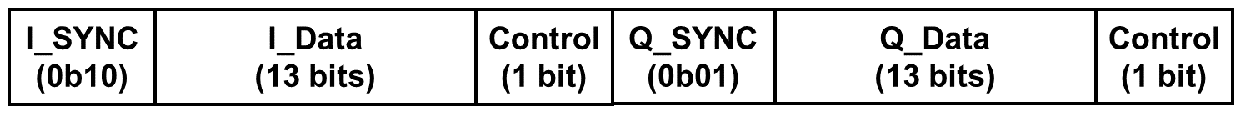}
	\vskip -0.15 in
	\caption{\footnotesize{{\bf I/Q Word Structure Used in I/Q Interface.}}}
	\label{fig:iq_structure}
	\vskip -0.2 in
\end{figure}

{\bf Receiving serial I/Q data.}
Our system communicates over LVDS to the FPGA in serial mode to transfer I/Q data with a physical interface consisting of 4 I/O lines, pairs of which are used to send data and clock signals. The radio outputs 32-bit serial data words at 4~Mwords/s using the format in Fig.~\ref{fig:iq_structure}. Each data word starts with the $I\_{SYNC}$ pattern which indicates the start of the $I$ sample which we use for synchronization. Next, it has 13 bits of $I\_{Data}$ followed by a control bit. The same format follows for $Q$, beginning with a synchronization pattern $Q\_{SYNC}$ and then 13 bits for $Q\_{Data}$ and the final control bit. The required 128~Mbps data rate is achieved using a 64~MHz clock provided by the radio operating at double data rate by sampling at both the rising and falling edges of the clock. We implement an I/Q deserializer on the FPGA to read the data which samples the input at both the rising and falling edges of the clock, uses the $I\_{SYNC}$ and $Q\_{SYNC}$ to detect the beginning of the data fields and loads the $I$ and $Q$ values into 13~bit registers for parallel processing.
 
{\bf Transmitting I/Q samples.} In TX mode we need to do the opposite of the above sequence to convert from the parallel representation on the FPGA to a serialized LVDS stream. To do this, we use the FPGA's onboard PLL to generate the 64~MHz clock signal. Next to create our double data rate output signal that varies on both the positive and negative edges of this clock signal using a dual-edge D flip-flop design~\cite{hildebrandt2011pseudo} resulting in the desired 128~Mbps data rate. We use this to generate the same I/Q word structure described above.

\subsubsection{Memory Interfaces}
After reading the raw data from the LVDS lines using the I/Q deserializer described above, we store the samples into a FIFO buffer implemented using the FPGA's embedded SRAM. We implement a simple memory controller to write data to the FIFO which generates the memory control signals and writes a full data word on each cycle. The embedded memory can run at rates significantly greater than 4~MHz meaning it is not a limiting factor for real time processing. The SRAM can buffer up to 126~kB. The data stored in the FIFO can then be sent to signal processing blocks to implement filters, cryptographic functions, etc. or to non-volatile flash memory. For flash memory, we use microSD cards which support two modes: native SD mode and standard SPI mode. In native SD mode, we use 4 parallel data lines to read/write data to/from the microSD card. This mode supports a higher data rate compared to the SPI mode which only supports 1 bit serial interface. However, we implement SPI mode since it supports the 104~Mbps data rate which we need to write data in real time. This allows us to re-use the same, simpler SPI block for multiple functions and save resources on the FPGA.

\subsubsection{RF, Control and Sensor Interfaces} 
The AT86RF215  provides differential RF signals for both 900~MHz and 2.4~GHz and has an integrated TX/RX switch for both. At 2.4~GHz, the differential signal is transformed to a single-ended output using the 2450FB15A050E~\cite{24balun} balun and fed to the SKY66112~\cite{SKY66112} front-end with the bypassable LNA and PA. Finally, after passing through a matching network, the 2.4~GHz signal is connected to an SMA output. 

On the 900~MHz side, the differential output of the AT86RF215 is connected to 0896BM15E0025E~\cite{900balun} to convert it to a single-ended output. This must be shared between the  backbone radio's two separate RF paths for transmit and receive and AT86RF215's 900~MHz single-ended signal. We choose between them using a ADG904~\cite{adg904} SP4T RF switch. The single port side is connected to the SE2435L~\cite{SE2435L} 900~MHz front-end which is similar to the 2.4~GHz front-end. The MCU communicates with the I/Q radio, backbone radio, FPGA and Flash memory through SPI which it uses to send commands for changing the frequency, selecting the outputs, etc. It also has control signals for FPGA programming, 900~MHz and 2.4~GHz front-end modules, RF switch and voltage regulators for active power control. {The I2C and SPI serial interfaces and analog to digital converter (ADC) inputs of the MCU are broken out on \name board to support both digital and analog sensors.}
% \textred{The MCU also supports I2C and SPI serial interfaces and analog to digital converter (ADC) inputs. Therefore, \name supports both digital and analog sensors.}

\subsection{Power Management Unit}
\label{sec:powermanagement}
Next, we present the design of our power management unit which seeks to maximize the system lifetime when running off of a 3.7~V Lithium battery. To enable long battery lifetimes we need to be able to duty-cycle our system and allow the MCU to toggle each of the above blocks ON and OFF when they are not in use. Further, different components have different supply voltage requirements and we wish to provide each one with the lowest voltage possible to minimize power usage. 

Ideally we would want separate controllable voltage regulators for each component in the system. However, having many different regulators with individual controls significantly increases the complexity, number of components, and price. Moreover, it complicates the PCB design by requiring many control signals and a multitude of power planes. Therefore, there exists a trade-off between the granularity of power control and the price/complexity of a design. We outline the supply voltages needed for each component and the power domain supporting it in Table~\ref{tab:voltages}. Below, we show how we group components to balance power and complexity. 

% Other than V1 which supplies the MCU, other voltage regulators would go to shut-down and wake-up. This means that V1 needs to have low quiesant current and others needs

\begin{table}[t]
    \centering
    \caption{\footnotesize{\bf Power Domains in \Name.}}
    \vskip -0.15 in
    \footnotesize{
    \begin{tabular}{|c|c|c|}
         \hline \rowcolor{lightgray}
         {\bf Component} & {\bf Voltage [V]} & {\bf Power Domain}\\
         \hline
         MCU & 1.8V & V1\\
         \hline
         FPGA & 1.1, 1.8, 2.5, Vlvds & V2, V3, V4, V5\\
         \hline
         I/Q Radio & 1.8< V5 <3.6 & V5\\
         \hline
         Backbone Radio & 1.8< V5 <3.6 & V5\\
         \hline
         sub-GHz PA & 3.5V & V6\\
         \hline
         2.4~GHz PA & 1.8, 3.0 & V3, V7\\
         \hline
         FLASH Memory & 1.8 & V3\\
         \hline
         Micro SD Memory& 3.0 & V7\\
         \hline
    \end{tabular}
    }
    \vskip -0.23 in
    \label{tab:voltages}
\end{table}
\squishlist
\item {\bf Power domain V1 (MCU).} Since the MCU is the central controller which implements power management, it needs to be powered at all times and therefore has its own power domain. To minimize its sleep current we need to use a voltage regulator with a low quiescent current. Although switching voltage regulators have higher conversion efficiency when active, they also have high quiescent currents so we instead select the TPS78218 linear regulator.

\item {\bf Power domains V2, V3, V4, V6 and V7.} These power domains provide power to blocks such as the FPGA, memory blocks, and PAs. Since these components can all be turned off when not operating, the voltage regulators for these domains should have low shut-down current during sleep and high efficiency when active. We therefore choose the TPS62240 which has a shutdown current of only 0.1~uA. It is highly efficient and is rated to support the current draw required by all components except the 900~MHz PA. To support this PA at its maximum output power we use the TPS62080 switching regulator which supports the required current. 

% We need a 1.8~V supply for the I/O banks and programming of the FPGA, the digital supply of the 2.4~GHz PA and flash memory which are all provided by V3 voltage domain. In addition, the FPGA I/O banks need a supply voltage of 2.5~V given by V4 in addition to V3.  Additionally, we need a 3~V voltage supply provided by V7 to power the microSD card and analog supply of the 2.4~GHz PA. The voltage regulator for V7 has the additional requirement of supporting currents as high as 120~mA when the 2.4~GHz PA is active. Similar to V2, we use three TPS62240 switching voltage regulators which can be set to output each of the required voltages for V3, V4 and V7 because of its low shut-down current, high efficiency and ability to provide up to 300~mA of current.

\item {\bf Power domain V5.} V5 is a shared power domain for I/Q radio, backbone LoRa radio and FPGA I/O bank. This power domain is initially set to 1.8V to minimize power consumption, however components such as the radio chips can require higher voltage to achieve maximum output power. Therefore, in addition to high efficiency and low shut-down current like the others, this domain should be programmable. To do this, we use Semtech SC195ULTRT~\cite{SC195ULTRT} which provides an adjustable output that can be set from 1.8~V to 3.6~V.
\squishend
% {\bf Voltage domain V6.} This voltage domain supplies the 900~MHz PA. This PA consumes as much as 550~mA current when transmitting at 30~dBm.  We use TPS62080~\cite{TPS62080} which is able to provide 1.2A. Note that,  we carefully design the ground and power plane on PCB to support enough current while minimizing heat dissipation issues. We also note that in normal operations the PA is turned off since IoT protocols do not such high transmit powers.

\subsection{Over-the-Air Programming protocol}
\label{sec:ota}
%Over-the-air updates require a protocol for interacting with the centralized access point (AP) to ensure each node receives the full bit stream, as well as minimizing transmission length to reduce network use time. %First, we explain how we prepare updates on the access point and handle transmissions to each node. Then, we explain the details of storing the update data and reprogramming the system.

{\bf OTA AP and MAC protocol.} To update a network of  \name devices, we use an AP with a LoRa radio to communicate with each device sequentially. In order to propagate updates throughout a testbed  or to specific \name nodes, we design a MAC layer for the LoRa PHY. %In order to avoid the chance of corrupted firmware due to packet loss during the update, and considering each update takes only XXX~s as shown in Fig.~\ref{xxx}, our access point updates each \name node individually.
We pre-program a timer on the MCU to periodically turn off the FPGA and switch from IQ radio mode to the backbone radio to listen for new firmware updates. If there is an update, the AP sends a programming request as a LoRa packet with specific device IDs indicating the nodes to be programmed along with the time they should wake up to receive the update. Upon processing this packet and detecting its ID, the \name node switches into update mode and sends a ready message to the AP at the scheduled time. Then, the AP transmits the firmware update as a series of LoRa packets with sequence numbers. Upon receiving each packet, the \name node checks the sequence number and CRC. For a correct packet it writes the data to its flash memory and transmits an ACK to indicate correct reception. In the case of failure no ACK is sent and the AP re-transmits the corrupted packet after a timeout. After sending all the firmware data, the AP sends a final packet indicating the end of firmware update which tells the \name node to reprogram itself and switch back to normal operation.

\begin{figure}[t]
    \centering
	\includegraphics[width=\linewidth]{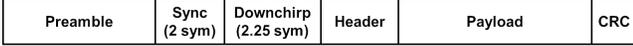}
	\vskip -0.1 in
	\caption{\footnotesize{{\bf LoRa Packet Structure.}}}
	\vskip -0.25in
	\label{fig:lora_packet}
\end{figure}

{\bf Compressing and decompressing the bitstream.} Our system compresses data to reduce update times, however this compression must be compatible with the resources available on \name. We choose the miniLZO compression algorithm~\cite{miniLZO}, which is a lightweight subset of the Lempel–Ziv–Oberhumer (LZO) algorithm. Our implementation of miniLZO only requires a memory allocation equal to the size of the uncompressed data. We perform compression on the AP. The compression ratio of bitstream file varies based on the content of the bitstream, and in the worst case the compressed file could have almost the same size of the original file. This would require a maximum memory allocation of 579~kB which we cannot afford on a low-cost MCU. Instead, we first divide the original update file into blocks of 30~kB that will fit in the MCU memory. Then we compress each block separately and transmit them to the \name node one by one. Considering the LoRa radio takes more power than the MCU, we immediately write the data to our dedicated programming flash memory using an SPI interface.

% \begin{figure}[t]
%     \centering
% 	\includegraphics[width=\linewidth]{./figs/block_diagram/lora_packet.eps}
% 	\vskip -0.1 in
% 	\caption{\footnotesize{{\bf LoRa Packet Structure.}}}
% 	\vskip -0.25in
% 	\label{fig:lora_packet}
% \end{figure}

\begin{figure}[t]
    \begin{subfigure}{\linewidth}
    \centering
        \includegraphics[width=1\linewidth]{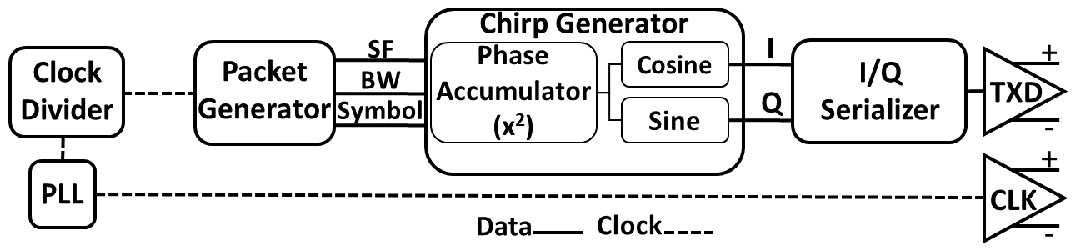}
        \vskip -0.1in
        \caption{\footnotesize{{\bf LoRa Modulator}}}
        \label{fig:lora_tx_block}
    \end{subfigure}
	\begin{subfigure}{\linewidth}
	    \centering
        \includegraphics[width=1\linewidth]{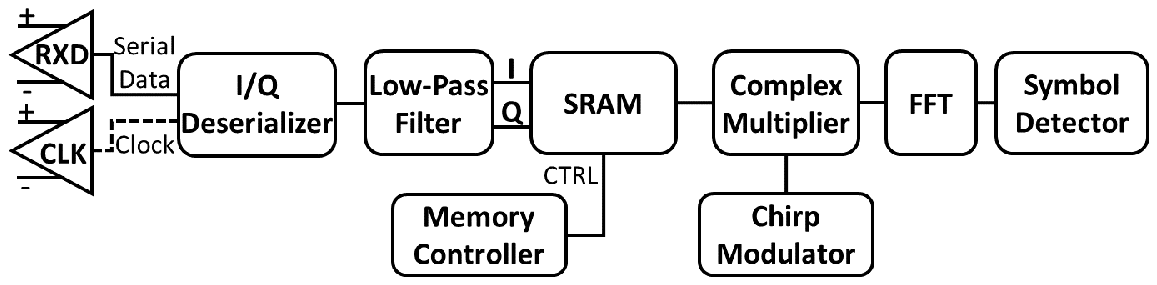}
        \vskip -0.1in
        \caption{\footnotesize{{\bf LoRa Demodulator}}}
        \label{fig:lora_rx_block}
    \end{subfigure}
    \vskip -0.1in
    \caption{\footnotesize{{\bf LoRa Implementation Block Diagrams.}}}
    \label{fig:lora_blocks}
    \vskip -0.2in
\end{figure}

After receiving all the data we turn off the LoRa radio and decompress data. First, we allocate memory on the MCU's SRAM equal to the block size and load a block of data from flash. Next, we perform decompression and write the data in the allocated SRAM memory. Finally, we write the decompressed data back to the flash beginning at the corresponding address of the programming boot file. We repeat these steps until we decompress the full firmware update.

{\bf Over-the-air FPGA programming.} After storing uncompressed programming data in flash memory, we program the FPGA. We use the MCU to set the FPGA into programming mode. When the FPGA switches to programming mode, it automatically reads { its firmware} directly from the flash memory using a 62~MHz quad SPI interface and programs itself. Reading from flash using quad SPI achieves programming times of 22~ms which is similar to FPGAs with embedded flash memory and results in minimal system down time. After programming is complete, it resumes operation and begins running the new firmware.

%Master quad SPI can achieve up to 62~Mbps data rate. Using quad SPI for programming is more necessary when FPGA wakes up from sleep mode. Since our FPGA is SRAM-based and does not include internal flash memory for programming, it requires external flash memory. When FPGA wakes up from sleep mode, it automatically program itself by reading programming file from external flash memory. In this case, using quad SPI reduces the programming time and it becomes similar to Flash-based FPGAs programming time.

\section{Case Studies: LoRa and BLE Beacons}
\subsection{LoRa Protocol with \name}
 We choose LoRa as it is gaining popularity for IoT solutions due to its long range capabilities. Since LoRa is a proprietary standard, we begin by describing the basics of its modulation and packet structure followed by the implementation details of our modulator, demodulator and MAC protocol.

{\bf LoRa Protocol Primer.} 
LoRa achieves long ranges by using Chirp Spread Spectrum (CSS) modulation. In CSS, data is modulated using linearly increasing frequency  upchirp symbol. Each upchirp symbol has two main features: Spreading Factor (SF) and Bandwidth (BW). SF determines the number of bits in each upchirp symbol~\cite{lorasigcomm17,netscatter,lorabackscatter} and BW is the difference between upper and lower frequency of the chirp which together with SF determines the length of an upchirp symbol. SF and BW  trade data rate for range. Data is modulated by $2^{SF}$ cyclic-shifts of an upchirp symbol. The starting point of the symbol in frequency domain, which is the cyclic shift of the upchirp symbol, determines its value~\cite{lora_basics}. LoRa uses SF values from 6 to 12 and BW values from 7.8125~KHz to 500~KHz to achieve PHY-layer rates of $\frac{BW}{2^{SF}}\times SF$. 

Fig.~\ref{fig:lora_packet} shows the LoRa packet structure which begins with a preamble of 10 zero symbols (upchirps with zero cyclic-shift). This is followed by the Sync field with two upchirp symbols. Next, a sequence of 2.25 downchirp symbols (chirp symbol with linearly decreasing frequency) indicate the beginning of the payload. The payload then consists of a sequence of upchirp symbols which encode a header, payload and CRC.

{\bf LoRa Modulator.} 
Fig.~\ref{fig:lora_tx_block} shows the block diagram of our LoRa modulator. We use our FPGA to implement a LoRa modulator in Verilog and stream data to AT86RF215 in I/Q mode. The modulator begins with the {\it Packet Generator} module which reads data either from FPGA memory for transmitting fixed packets or from the MCU, as well as LoRa configuration parameters such as SF, coding and BW. This module determines each symbol value and its corresponding cyclic-shift. Next, the {\it Packet Generator} sends these parameters along with the symbol values to the {\it Chirp Generator} module, which generates the I/Q samples of each chirp symbol in the packet using a squared phase accumulator and two lookup tables for $Sin$ and $Cos$ function~\cite{lorabackscatter}. We then feed these I/Q samples into {I/Q Serializer}  to stream them over the LVDS interface to the I/Q radio. We generate 64~MHz transmission clock using internal PLL of the FPGA.

\begin{figure}[t!]
    \centering
    \includegraphics[width=0.75\linewidth]{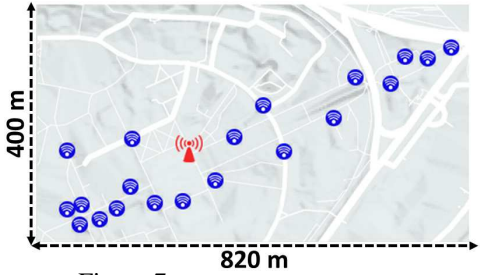}
    \vskip -0.15in
    \caption{\footnotesize{{\bf Evaluation Testbed Map.}}}
    \label{fig:ota_map}
    \vskip -0.2in
\end{figure}

{\bf LoRa Demodulator. }
Fig.~\ref{fig:lora_rx_block} shows the block diagram of our LoRa demodulator. It begins by reading data from the I/Q radio into the {\it I/Q Deserializer} module on the FPGA which converts the serial I/Q stream to parallel I/Q for further signal processing. Next, we run the data through a 14 tap FIR low-pass filter to suppress high frequency noise and interference. We store the filtered samples in a buffer implemented using the FPGA's memory blocks. To decode the data, we use the {\it Chirp Generator} module from the {\it LoRa Modulator} described above to generate a baseline upchirp/downchirp symbol, and then we multiply that with the received chirp symbol using our {\it Complex Multiplier} unit. The output of the multiplication then goes to an FFT block implemented using a standard IP core from Lattice. Finally the {\it Symbol Detector} scans the output of the FFT for peaks and records the frequency of the peak to determine the symbol value. To detect chirp type (upchirp/downchirp), we multiply each chirp symbol with both an upchirp and downchirp and then compare the amplitudes of their FFT peaks. The higher peak in the FFT shows higher correlation which indicates the chirp type. 

%\begin{figure}[t]
%    \centering
%	\includegraphics[width=\linewidth]{./figs/block_diagram/ble_tx_block.eps}
%	\vskip -0.15 in
%	\caption{\footnotesize{{\bf BLE Implementation Block Diagram}}}
%	\label{fig:ble_tx_block}
%	\vskip -0.2 in
%\end{figure}

% \begin{figure}[t!]
%     \centering
%     \includegraphics[width=0.75\linewidth]{figs/10_ota_exp/map.eps}
%     \vskip -0.15in
%     \caption{\footnotesize{{\bf Evaluation Testbed Map.}}}
%     \label{fig:ota_map}
%     \vskip -0.22in
% \end{figure}

{\bf LoRa MAC Layer.} 
To demonstrate that our LoRa implementation on \name is compatible with existing LoRa networks such as the LoRa Alliance's~\cite{loraalliance} {The Things Network} (TTN)~\cite{thingsnetwork}, we adopt their LoRa MAC design from TTN's Arduino libraries~\cite{ttnarduino} and implement it on \name's MCU. TTN uses two methods for device association; Over-the-air activation (OTAA) and activation by personalization (ABP). {In OTAA, each node performs a join-procedure during which a dynamic device address is assigned to a node. However, in ABP we can hard-code the device address in the device which makes it simpler since the node skips the join procedure. Our platform can support both OTAA and ABP methods.}

\subsection{BLE Beacons with \name}
To demonstrate \name's 2.4~GHz capabilities we implement Bluetooth beacons which are commonly used by IoT devices.

{\bf BLE Beacon Primer.} 
We implement non-connectable BLE advertisements (ADV\_NON\_CONN\_IND) which are broadcast packets used for beacons. These packets allow a low power device to broadcast its data to any listening receiver within range without the power overhead of exchanging packets to  setup a connection. These packets have a bit rate of 1~Mbps in Bluetooth 4.0 or up to 2~Mbps in Bluetooth 5.0 and are generated using GFSK with a modulation index of 0.45-0.55. The GFSK modulation is  binary frequency shift keying (BFSK) with the addition of a Gaussian filter to the square wave pulses to reduce the spectral width.

\begin{figure}[t]
    \centering
    \includegraphics[width=1\linewidth]{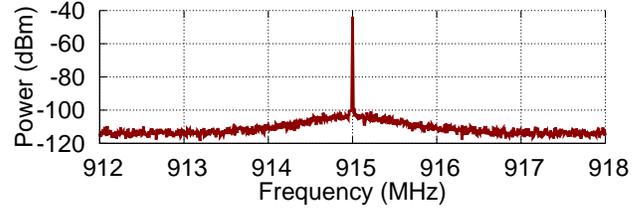}
    \vskip -0.15in
    \caption{\footnotesize{{\bf \Name Single-Tone Frequency Spectrum.}}}
	\label{fig:spectrum}
    \vskip -0.11in
\end{figure}

\begin{table}[t!]
    \centering
    \caption{\footnotesize{\bf Different Operation Timing for \Name.}}
    \vskip -0.15 in
    \footnotesize{
    \begin{tabular}{|C{4cm}|C{2cm}|}
         \hline \rowcolor{lightgray}
         {\bf Operation} & {\bf Duration (ms)}\\
         \hline
         Sleep to Radio Operation & 22\\ 
         \hline
         Radio Setup & 1.2\\ %Reg_wakeup(500)+wake_trxoff+traxoff_txprep
         \hline
         TX to RX & 0.045\\     %TX to TRXprep(33us) + TRXprep to RX(200ns) + SPI (10us)
         \hline
         RX to TX & 0.011\\       %RX to TRXprep(200ns) + TRXprep to TX(200ns) + SPI (10us)
         \hline
         Frequency Switch & 0.220\\ %we measured this
         \hline
    \end{tabular}
    }
    \vskip -0.2in
    \label{tab:wakeup}
\end{table}

{\bf Generating a BLE Packet.} 
Bluetooth advertisements consist of 6-37 octets, beginning with fixed preamble and access address fields indicating the packet type set to 0xAA and 0x8E89BED6 respectively. This is followed by the packet data unit (PDU) beginning with a 2 byte length field and followed by a manufacturer specific advertisement address and data. The final 3 bytes of the packet consist of a CRC generated using a 24-bit linear feedback shift register (LFSR) with the polynomial $x^{24} + x^{10} + x^{9} + x^{6} + x^{4} + x^{3} + x + 1$. The LFSR is set to a starting state of 0x555555 and the PDU is input LSB first. The final LFSR state after inputting the PDU becomes the CRC. Data whitening is then performed over the PDU and CRC fields to eliminate long strings of zeros or ones within a packet. This is also done using a 7-bit LFSR with polynomial $x^7 + x^4 + 1$. The LFSR is initialized with the lower 7 bits of the channel number the packet will be transmitted on, and each byte is input LSB first. We implement both these blocks in Verilog on the FPGA.

{\bf Packet  Transmission and MAC Protocol.} From this bitstream, we need to generate the I/Q samples to feed to the I/Q radio. First, we upsample and apply a Gaussian filter to the bitstream. This gives us the desired changes in frequency which we integrate to get the phase. We then feed the phase to sine and cosine functions to get the final I and Q samples, which are passed to I/Q serializer and sent to the I/Q radio. BLE divides the 2.4~GHz band into channels, each spaced 2~MHz apart, but BLE beacons are only transmitted on three advertising channels without carrier sense, typically in sequential order separated by a few hundred microseconds. This sequence is re-transmitted every advertising interval~\cite{ble_spec}.

% \subsection{Orthogonal LoRa Demodulation}
% In this section, we demonstrate that we can demodulate multiple orthogonal chirp symbols on \name. 

% \subsubsection{LoRa orthogonal transmissions.}
% LoRa protocol supports orthogonal transmissions~\cite{lora_basics} which can occupy the same frequency channel and does not interference with each other. LoRa uses orthogonal transmissions to increase network capacity. For a chirp with spreading factor of $SF$ and bandwith of $BW$, we define chirp slope as the bandwidth occupation of the chirp over time and it is qual to $\frac{BW^{2}}{2^{SF}}$~\cite{netscatter}. Two chirp symbols are orthogonal when they have different chirp slope. Semtech chipset such as SX1308~\cite{sx1308} provides programmable parallel demodulation for different LoRa configurations. Here, we show that we can also demodulate different multiple LoRa transmissions at the same time using \name.

% \subsubsection{Orthogonal demodulation}
% We target to demodulate two upchirp symbols which has two configurations; $SF1, BW1$ and $SF2, BW2$. First, we generate two downchirp symbols with the same configurations. We perform correlation on each separately by multiplying each upchirp symbol to the corresponding downchirp symbol. After correlation, we use FFT operation to detect the maximum peak power of each upchirp symbol. We perform all operations on two parallel pip-lines to demodulate two upchirp symbols concurrently. 

\section{Evaluation}
We deploy a testbed of 20 \name devices across  our institution's campus as shown in Fig.~\ref{fig:ota_map} (details removed for anonymity).
To see if \name meets the requirements for IoT endpoint devices, we characterize its power, computational resource usage, delays and cost when operating in different modes and running different protocols. 

%We begin with benchmarks and specs detailing radio performance and power consumption followed by cost estimates. We then present the results with our two case studies. Following this, we evaluate the performance of our over the air update system.

\vspace{-0.1 in}
\subsection{Benchmarks and Specifications}
{\bf Sleep mode power.} Many IoT nodes perform short, simple tasks allowing them to be heavily duty cycled which allows them to achieve battery lifetimes of years. We design \name with this critical need in mind such that the MCU can actively toggle on and off power consuming components such as the radio, PAs, and FPGA to enter a low power sleep mode.

We do this by first turning off the the I/Q transceiver and LoRa radios. To reduce the static power consumption of the FPGA, we shut it down by disabling the voltage regulators that provide power to its I/O banks and core voltage. Similarly, we also turn off the PAs. Finally, we put the MCU in sleep mode LPM3 running only a wakeup timer. The measured total system sleep power in this mode was 30~uW.

The low sleep power allows for significant power savings, but also introduces latency. Table~\ref{tab:wakeup} shows the time required to wake up from sleep mode until the radio is active. Because we can perform the I/Q radio setup in parallel with booting the FPGA, the total wakeup time for RX and TX is 22~ms. The I/Q radio setup takes 1.2~ms, so the wakeup time is dominated by booting up the FPGA which itself takes 22~ms. We compare this to a SmartSense Temperature sensor~\cite{smartthings_temp} and find that \name has only a 4x longer wakeup time even though it requires programming unlike commercial products that use a custom single protocol radio. Additionally many IoT devices operate at low duty cycles waiting in sleep mode for seconds or more making \name's wakeup latency insignificant.
%A wakeup time of 21~ms is however within the expected wakeup times of IoT devices including XXX and XXX~\cite{XXX}. To fully illustrate the wakeup process we demonstrate a typical IoT use case. Specifically, we measure the power consumption of our platform as it wakes up from sleep mode, measures a sensor value, and transmits a packet. %Fig.~\ref{fig:wakeup_curve} shows the power consumption versus time and results in an average power of xx for an xx\% duty cycle. This means that \name can achieve a battery life of xx using an xx battery.

{\bf Switching delays.} We also measure the switching delays for different operations on the I/Q radio as this is an important parameter for meeting MAC and protocol timing requirements.  Table~\ref{tab:wakeup} shows that it takes 45~$\mu$s and 11~$\mu$s to switch from TX to RX mode and RX to TX mode respectively. As we see later, this is sufficient to meet the timing requirements of IoT packet ACKs and MAC protocols. Further, the delay for switching between different frequencies is only 220~us. To measure this number, we switch between 2.402~GHz, 2.426~GHz and 2.480~GHz. This switching delay is again sufficient to meet the requirements of frequency hopping during Bluetooth advertising.

\begin{table}[t]
\centering
\caption{\footnotesize{\bf \Name Cost Breakdown for 1000 Units.}}
\vskip -0.15in
\footnotesize{
    % \begin{tabular}{|C{3cm}|C{2.3cm}|C{1.5cm}|}
    \begin{tabular}{|C{3cm}|C{2cm}|C{1.5cm}|}
    \hline
    \cline{1-3}\rowcolor{lightgray}
    \multicolumn{2}{|c|}{\bf Components}    &  {\bf Price}      \\\hline
    \multirow{2}{*}{DSP} &    FPGA                &  \$8.69           \\\cline{2-3}
                   &    Oscillator          & \$0.9             \\\hline
    \multirow{4}{*}{IQ Front-End}                 & Radio &   \$5.08  \\\cline{2-3}
                                            & Crystal       & \$0.53    \\\cline{2-3}
                                            & 2.4~GHz Balun & \$0.36    \\\cline{2-3}
                                            & Sub-GHz Balun & \$0.3     \\\hline
    \multirow{3}{*}{Backbone}                     & Radio &   \$4.5     \\\cline{2-3}
                                            & Crystal & \$0.4           \\\cline{2-3}
                                        & Flash Memory      & \$1.6     \\\hline
    \multirow{2}{*}{MAC}                & MCU               & \$3.89    \\\cline{2-3}
                                        & Crystals          & \$0.68    \\\hline
    \multirow{3}{*}{RF}                 & Switch            & \$3.14    \\\cline{2-3}
                                        & Sub-GHz PA        & \$1.54    \\\cline{2-3}
                                        & 2.4~GHz PA        & \$1.72    \\\hline
    \multirow{1}{*}{Power Management}   & Regulators        & \$3.7     \\\hline
    \multirow{1}{*}{Supporting Components}      & --        & \$4.5     \\\hline
    \multirow{2}{*}{Production}         & Fabrication~\cite{pcbminions} & \$3   \\\cline{2-3}
                                        & Assembly~\cite{pcbminions}    & \$10  \\\hline
    \multirow{1}{*}{\bf Total}          & --  & {\bf \$54.53}    \\\hline
    \end{tabular}
}
\vskip -0.2in
\label{tab:cost}
\end{table}

\begin{figure}[t]
    \centering
    \includegraphics[width=1\linewidth]{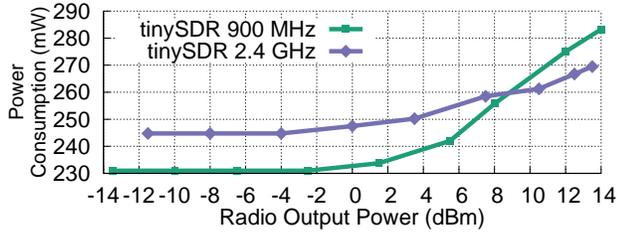}
    \vskip -0.15in
    \caption{\footnotesize{{\bf Single-Tone Transmitter Power Consumption.} We show the total power consumption of \name including I/Q radio, FPGA,  MCU  and regulators at different transmitter output power. This is 15-16 times lower power consumption than the USRP E310 embedded SDR.}}
	\label{fig:bench_tx_power}
    \vskip -0.22in
\end{figure}

{\bf Transmitter performance.} First,  we implement a single-tone modulator on the FPGA that generates the appropriate I/Q samples and streams them over LVDS to the radio. We connect the output to an MDO4104b-6~\cite{mdo4104b} spectrum analyzer and observe a single tone, shown in Fig.~\ref{fig:spectrum}, with no unexpected harmonics introduced by the modulator.

Next we measure the end-to-end DC power consumption of our system including the I/Q radio, FPGA, MCU and regulators to see how it scales with RF output power. We vary our radio output power while transmitting a single tone and use a Fluke 287 multimeter to measure its DC power draw. Fig.~\ref{fig:bench_tx_power} shows the power consumption of \name for 900~MHz and 2.4~GHz operation. Interestingly, we observe the DC power is constant at low RF power but increases as expected beyond some RF power level. \Name consumes 231~mW when transmitting at 0~dBm, and for comparison the end-to-end power consumption of the USRP E310 is 16x higher under the same conditions. Similarly \name consumes 283~mW at its 14~dBm setting while the USRP E310 is 15x higher.

{\bf Cost.} We also analyze the cost which is an important practical consideration for real world deployment at scale. Table~\ref{tab:cost} shows a detailed breakdown of cost including each component as well as PCB fabrication and assembly based on quotes for 1000 units~\cite{pcbminions}, where the overall cost is  around  \$55.

\subsection{Evaluating the Case Studies}
%To better evaluate the performance of \name, we evaluate our two protocols: LoRa and BLE beacons.

{\bf LoRa using \name.} 
We evaluate various different components of \name\ using LoRa as a case study.

\begin{figure}[t]
    \centering
    \includegraphics[width=1\linewidth]{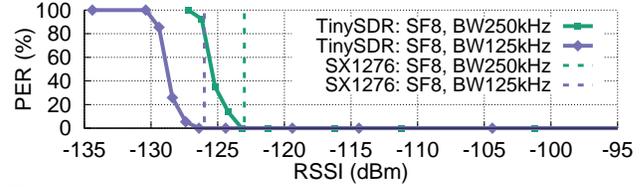}
    \vskip -0.15in
    \caption{\footnotesize{{\bf LoRa Modulator Evaluation.} We evaluate our LoRa modulator in comparison with Semtech LoRa chip.}}
	\label{fig:bench_lora_tx_per}
    \vskip -0.15in
\end{figure}

{\it LoRa modulator.} 
 To evaluate this, we use our LoRa modulator to generate packets with three byte payloads using a spreading factor of $SF=8$ and bandwidths of 250~kHz and 125~kHz which we transmit at -13~dBm. We receive the output of \name on a  Semtech SX1276 LoRa transceiver~\cite{sx1257} which we use to measure the packet error rate (PER) versus RSSI and plot the results in Fig.~\ref{fig:bench_lora_tx_per}. We compare our LoRa modulator to transmissions from an SX1276 LoRa transceiver. The plots show that we can achieve a comparable sensitivity of -126~dBm which is the LoRa sensitivity for $SF=8$ and $BW=125kHz$ configuration. This is true for both configurations, which shows that our low-power SDR can meet the sensitivity requirement of LPWAN IoT protocols.

\begin{figure}[t]
    \centering
    \includegraphics[width=1\linewidth]{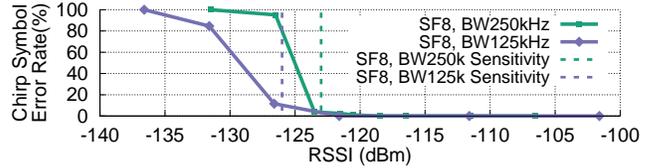}
    \vskip -0.15in
    \caption{\footnotesize{{\bf LoRa Demodulator Evaluation.} We evaluate our LoRa demodulator by demodulating chirp symbols at different RSSI.}}
	\label{fig:bench_lora_rx_per}
    \vskip -0.2in
\end{figure}

{\it LoRa demodulator.} Next we evaluate our LoRa demodulator on \name. To test this, we use transmissions from a Semtech SX1276 LoRa transceiver and use \name\ to receive these transmissions. The LoRa transceiver transmits packets with two configurations using a spreading factor of 8 and bandwidths of 250~kHz and 125~kHz. We record the received RF signals in the FPGA memory and run them through our demodulator to compute a chirp symbol error rate. {Note that the Semtech LoRa transceiver does not give access to symbol error rate but  since  we have  access to I/Q samples, we can  compute it on our platform.} We plot the results in Fig.~\ref{fig:bench_lora_rx_per} as a function of the LoRa RSSI values. Our LoRa demodulator can demodulate chirp symbols down to -126~dBm which is LoRa protocol sensitivity at $SF=8$ and $BW=125kHz$. Both the LoRa modulator  and demodulator run in real-time. 

\begin{table}[t]
    \centering
    % \begin{minipage}[t]{\linewidth}
        \caption{\footnotesize{{\bf FPGA Utilization for LoRa Protocol.}}}
        \vskip -0.15 in
        \footnotesize{
        \begin{tabular}{|C{1cm}|C{2.5cm}|C{2.5cm}|}
             \hline \rowcolor{lightgray}
             {\bf SF} & {\bf LoRa TX (LUT)} & {\bf LoRa RX (LUT)}\\
             \hline
             6 & 976 (4\%) & 2656 (10\%)\\
             \hline
             7 & 976 (4\%) & 2670 (10\%)\\
             \hline
             8 & 976 (4\%) & 2700 (11\%)\\
             \hline
             9 & 976 (4\%) & 2742 (11\%)\\
             \hline
             10 & 976 (4\%) & 2786 (11\%)\\
             \hline
             11 & 976 (4\%) & 2794 (11\%)\\
             \hline
             12 & 976 (4\%) & 2818 (11\%)\\
             \hline
        \end{tabular}
        }
        \label{tab:lora_utilization}
        \vskip -0.2in
    % \end{minipage}
    % \hspace{0.2in}
    % \begin{minipage}[t]{0.45\linewidth}
    %     \caption{\footnotesize{\bf Power consumption of different protocols.}}
    %     \vskip -0.15 in
    %     \footnotesize{
    %     \begin{tabular}{|C{1.5cm}|C{1.4cm}|}
    %          \hline \rowcolor{lightgray}
    %          {\bf Protocol} & {\bf Power (mW)}\\
    %          \hline
    %          LoRa RX & 186\\    %(112mW + 30mA*1.8+4mW MCU)*(Reg. Efficiency = 1.1)
    %          \hline
    %          LoRa TX & 287\\    %(95mW + 65mA*2.5+4mW MCU)*(Reg. Efficiency = 1.1)
    %          \hline
    %          BLE Beacon & 263\\ %(101mW + 51mA*2.5+4mW MCU)*(Reg. Efficiency = 1.1)
    %          \hline
    %     \end{tabular}
    %     }
    %     \vskip -0.2in
    %     \label{tab:protocol_pow}
    % \end{minipage}
\end{table}

{\it Resource allocation.} Next, we evaluate the resource utilization of our LoRa PHY implementation on the FPGA. Table~\ref{tab:lora_utilization} shows the size for implementing the modulator and demodulator on our FPGA performing using different SFs. Our LoRa modulator supports all LoRa configurations with different SF with no additional cost. However, in the LoRa demodulator, we need FFT blocks with different sizes to support different SF configurations. This table shows that our FPGA has sufficient resources to support multiple configurations of LoRa and still leave space for other custom operations. 

% ??? There should be something to evaluate it? XXXX Delay for ACK? or scheduling? XXXX CDF?
%(32+15)/256=18%
%5/65
{\it LoRa MAC.} We implement the LoRa MAC based on TTN's Arduino libraries~\cite{ttnarduino}. TTN protocol together with control for the I/Q radio, backbone radio, FPGA, PMU and decompression algorithm for OTA take only 18\% of MCU resources. Also, as shown in Table~\ref{tab:wakeup}, our timings are well within the requirements for LoRaWAN specifications~\cite{loraalliance}. 

% XXXWhat is the system power consumption of transmitting a Lora packet? What about receiving? XXXX What are the individual power consumption?XXX
We also measure the power consumption of our platform for LoRa packet transmission and reception. LoRa packet transmission with $SF=9$ and $BW=500~kHz$ and radio output power of 14~dBm consumes a total power of 287~mW from which 179~mW is for the radio and the rest is for the FPGA and MCU. LoRa packet reception consumes 186~mW with radio taking 59~mW.
% In comparison with a non-SDR IoT LoRa chipset such as SX1276, we consume only 3 times more power, while providing full SDR capabilities. SX1276 packet reception takes around 20~mW.

\begin{figure}[t]
    \centering
    \includegraphics[width=1\linewidth]{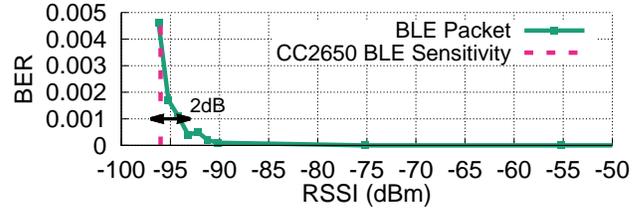}
    \vskip -0.15in
    \caption{\footnotesize{{\bf BLE evaluation.} BLE beacons at different power levels.}}
	\label{fig:bench_ble_tx_ber}
    \vskip -0.2in
\end{figure}

{\bf BLE using \name.} Next, we evaluate \name\ using BLE beacons as a case study. First, we measure the impact of our BLE beacons transmitted from \name\ using the TI CC2650~\cite{cc2650} BLE chip as a receiver. We do this by configuring \name to transmit BLE beacons at a rate of 1 packet per second.  We transmit 100 packets and set the CC2650 BLE chip to report bit error rate (BER). Fig.~\ref{fig:bench_ble_tx_ber} shows the BER as a function of the received RSSI as reported by the CC2650 BLE chip. The plot shows that we achieve a sensitivity of -94~dBm. This is  within 2~dB of the CC2650 BLE chipset's sensitivity, defined by a BER threshold of $10^{-3}$.

\begin{figure}[t]
    \centering
    \includegraphics[width=1\linewidth]{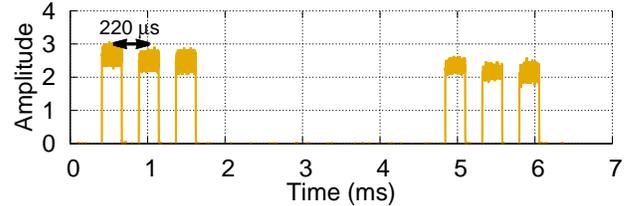}
    \vskip -0.15in
    \caption{\footnotesize{{\bf BLE Beacons Signal.} We show BLE beacon transmissions on three advertising channels from \name using an envelope detector.}}
	\label{fig:ble_timing}
    \vskip -0.2in
\end{figure}

Next we evaluate the latency of our BLE implementation as BLE beacons are typically transmitted in sequence by hopping between three different advertising channels. We measure the minimum time \name takes to switch between these frequencies by connecting its output to a 2.4~GHz envelope detector and using an MDO4104B-6 oscilloscope to measure the time delay between transmissions. Fig.~\ref{fig:ble_timing} plots the envelope of three BLE beacons in the time-domain transmitted on the different advertising channels and shows that our system can transmit packets with as little as 220~us delay between beacons. The corresponding result when a iPhone 8 transmits beacons is 350~us. Finally, generating BLE beacons requires only 3\% of the FPGA resources on the \name and it could run for over 2 years on a 1000~mAh battery when transmitting once per second.

\subsection{Over-the-Air Programming}
 An effective OTA programming system should both minimize use of system resources such as power as well as network downtime. Considering the time to reprogram the FPGA and microcontroller from flash is fixed, the downtime for programming a node depends on the amount of data sent and the throughput which varies with SNR. 

Raw programming files for our FPGA are 579~kB, however  we compress our data using miniLZO. While the exact compression ratio depends on FPGA utilization, our LoRa program compresses to 99~kB and BLE to 40~kB. Our microcontroller programs for both LoRa and BLE are approximately 78~kB and are both compressed to 24~kB. When dividing the files into packets, we would ideally minimize the preamble length and maximize packet length to reduce overhead, however long packets with short preambles lead to higher PER. We choose a preamble of 8 chirps and packets of 60~B which we find balances the trade-off of protocol overhead versus range in our experiments. 

\begin{figure}[t]
    \centering
    \includegraphics[width=1\linewidth]{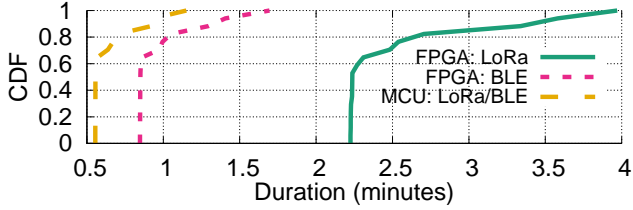}
    \vskip -0.15in
    \caption{\footnotesize{{\bf OTA Programming Time.} We show CDF of OTA programming time for programming LoRa and BLE implementations on \name.}}
	\label{fig:ota_exp}
    \vskip -0.2in
\end{figure}

To see the impact on a real deployment, we evaluate the time required to program \name nodes in our 20 device testbed shown in Fig.~\ref{fig:ota_map}. We set up a LoRa transceiver configured with $SF=8$, $BW=500~kHz$ and $CodingRate=6$ connected to a patch antenna transmitting at 14~dBm as an AP and measure the time it takes to program the \name devices at each location, according to our protocol. We transmit the compressed FPGA and MCU programming data for LoRa and BLE and plot the results as a CDF in Fig.~\ref{fig:ota_exp}. The plots show that the LoRa FPGA requires an average programming time of 150~s while BLE, FPGA, and MCU require 59~s and 39~s respectively due to their smaller file size. { Decompressing these received files only takes a maximum of 450~ms.}

Our OTA programming system components, backbone radio and MCU, consume an average energy of 6144~mJ for receiving a LoRa FPGA update and 2342~mJ for a BLE FPGA update when using 14~dBm output power. Using a 1000~mAh LiPo battery, we could OTA program each \name node with LoRa 2100 times and BLE 5600 times. Assuming OTA programming of once per day, the average power consumption would be 71~uW and 27~uW respectively for LoRa and BLE.

% \begin{figure}[t]
%     \begin{subfigure}{0.49\linewidth}
%     \centering
%         \includegraphics[width=1\linewidth]{figs/10_ota_exp/10_ota_time_cdf.eps}
%         \vskip -0.1in
%         \caption{\footnotesize{{\bf Programming Time}}}
%         \label{fig:ota_exp}
%     \end{subfigure}
% 	\begin{subfigure}{0.48\linewidth}
% 	    \centering
%         \includegraphics[width=0.59\linewidth]{figs/10_ota_exp/map.eps}
%         \vskip -0.1in
%         \caption{\footnotesize{{\bf Evaluation Map}}}
%         \label{fig:ota_map}
%     \end{subfigure}
%     \vskip -0.1in
%     \caption{\footnotesize{{\bf OTA Programming Evaluation.} We show CDF of OTA programming time for programming LoRa and BLE implementations.}}
%     \label{fig:ota_eval}
% \end{figure}

\vspace{-0.1 in}
\section{Research Study: Concurrent Reception}
An SDR designed for IoT endpoints that can provide I/Q transmission and reception capability opens up opportunities for addressing multiple research questions in IoT networks.

In this section, we focus on the following question: Can a {\it low-power IoT endpoint} device decode multiple concurrent LoRa transmissions at the same time? LoRa supports long range communication for IoT devices and is gaining popularity as a low-power wide area networking (LPWAN) standard. Supporting long ranges  introduces new challenges since it increases the probability of collisions  in large scale city-wide deployments. While recent works~\cite{lorasigcomm17,netscatter} have explored the feasibility of enabling concurrent LoRa transmissions, they have been designed for decoding on a gateway-style USRP device. In fact, most concurrent transmission techniques in our community~\cite{zigzag, anc, lorasigcomm17} have been prototyped on USRPs and it is unclear if a low-power IoT endpoint device can decode concurrent transmissions in real-time within its stringent power and resource constraints. TinySDR enables us to explore such questions and design MAC protocols for decoding concurrent transmissions on IoT endpoints. 

{\bf Using orthogonal LoRa codes.} Here we explore a specific way of enabling concurrent transmissions in LoRa: using orthogonal codes. Specifically, to allow multiple LoRa nodes to communicate at the same time, we exploit LoRa's support for orthogonal transmissions~\cite{lora_basics} which can occupy the same frequency channel without interfering with each other. Two chirp symbols are orthogonal when they have a different chirp slope. For a chirp with a spreading factor of $SF$ and bandwidth of $BW$, the chirp slope is given by: $\frac{BW^{2}}{2^{SF}}$~\cite{netscatter}.

{\bf Decoding concurrent transmissions on \name.} In order to receive concurrent LoRa transmissions, \name\ must be able to demodulate LoRa upchirp symbols with different slopes. Suppose we have two LoRa transmissions that use different spreading factor and bandwidth configurations: $SF_1, BW_1$ and $SF_2, BW_2$. To decode them concurrently, we implement decoders similar to Fig.~\ref{fig:lora_rx_block} for each chirp configuration in parallel on our FPGA.  Specifically, we first generate a corresponding downchirp symbol for each configuration. We then correlate the received signals with their corresponding downchirp symbols using time domain multiplication. After correlation, we take the appropriate length FFT of the result.

{\bf Evaluation.} We evaluate three key aspects of our design: 1) the  platform's effectiveness in decoding concurrent transmissions across a range of RSSI values, 2) the power consumption at the endpoint device while decoding concurrent transmissions and 3) the computational resources required. 

We use two SX1276 LoRa transceivers as our transmitters and set them to transmit continuously at two different settings: they both use a spreading factor of $SF=8$ but have two different bandwidth setups, $BW_1=125 kHz$ and $BW_2=250 kHz$. We set the two  to send random chirp symbols. The \name platform decodes these two concurrent transmissions and computes the chirp symbol error rate for each  transmission. We evaluate two scenarios: 1) when the two transmitters have a similar power level at the receiver, 2) fix the power of one of the transmitters and increase the power of the other one.

% \begin{figure}[t]
%     \centering
%     \includegraphics[width=1\linewidth]{figs/8_lora_orthogonal_rx/lora_multiple.eps}
%     \vskip -0.15in
%     \caption{\footnotesize{{\bf LoRa Orthogonal Demodulator Evaluation.}}}
% 	\label{fig:lora_multiple_rx}
%     \vskip -0.15in
% \end{figure}

\begin{figure}[t]
    \begin{subfigure}{1\linewidth}
    \centering
        \includegraphics[width=1\linewidth]{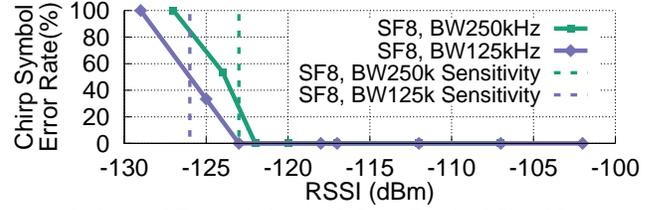}
        \vskip -0.1in
        \caption{\footnotesize{{\bf Orthogonal Transmissions with Same Received Signal Power.}}}
        \label{fig:lora_multiple_rx_same}
    \end{subfigure}
	\begin{subfigure}{1\linewidth}
	    \centering
        \includegraphics[width=1\linewidth]{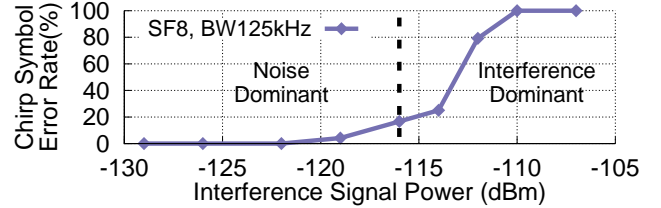}
        \vskip -0.1in
        \caption{\footnotesize{{\bf Orthogonal Transmissions with Different Received Signal Power}. We set the power of LoRa transmission with $BW_1=125kHz$ to -123~dBm and increase the power of the other one.}}
        \label{fig:lora_multiple_rx_diff}
    \end{subfigure}
    \vskip -0.1in
    \caption{\footnotesize{{\bf Orthogonal LoRa Demodulation Evaluation}}}
    \label{fig:lora_multiple_rx}
    \vskip -0.2in
\end{figure}

Fig.~\ref{fig:lora_multiple_rx_same} shows the results when the two transmissions have similar power at the receiver. We lose around 2~dB and 0.5~dB sensitivity for concurrent demodulation of LoRa configurations with $BW_1=125kHz$ and $BW_2=250kHz$. This is because while in theory the two chirps are orthogonal, in practice, the chirps are created in the digital domain with discrete frequency steps which introduces some non-orthogonality.

Fig.~\ref{fig:lora_multiple_rx_diff} shows the results when the first  LoRa transmitter $BW_1=125kHz$ is received  near its sensitivity  of   -123~dBm and and the second LoRa transmitter changes its power. Here, the  chirp symbol error rate is affected when the other transmission's power is higher than -116~dBm. When two concurrent transmissions are present, one acts as an interferer when decoding the other. The combined power of noise and the interferer, $P_{I,N}$, determines the error rate. When sweeping the power of interferer, at first the $P_{I,N}$ is dominated by noise and we should not see much effect on error rate. Then at some point the power of them would be equal which results in a 3~dB increase of $P_{I,N}$ and hence 3~dB sensitivity loss and afterwards the the error rate is determined by the interferer power. This demonstrates the need for power  control for  concurrent transmissions to be received on IoT endpoints.

Our parallel demodulation implementation, uses only 17\% of the FPGAs resources. This concurrent demodulation implementation consumes 207~mW. Note that Semtech gateway solutions such as SX1308~\cite{sx1308} can receive multiple transmissions. But,  to the best of our knowledge we are the first to show that concurrent LoRa transmissions can be decoded on a IoT endpoint while meeting its power and computational requirements, which is difficult to do without \name.

\section{Conclusion and Research Opportunities}
This paper presents the first SDR  platform specifically tailored to the needs of  IoT endpoints that can be used for large scale IoT network deployments. The goal of \name  is to provide a platform that can catalyze research  in  IoT networks. 

{\bf Research on PHY/MAC protocols.} \Name  presents an opportunity for researchers to avoid the time consuming endeavor of building their own custom hardware and instead focus on PHY/MAC protocol innovations across the stack: What is the trade-off between packet length and overall throughput? Are there benefits of rate adaptation? What about concurrent transmissions from IoT devices? One can also create multi-hop IoT PHY/MAC innovations, which have not been explored well given the lack of a flexible platform.

{\bf Research on IoT localization.} \Name could also be used to build localization systems as it gives access to I/Q signals and therefore phase  across 2.4~GHz and 900~MHz bands, which forms the basis for many localization algorithms\cite{sensys18}. One can also explore distributed localization solutions that combine the phase information across a distributed set of sensors to create a large MIMO sensing system. 

{\bf Machine learning on IoT devices.} The FPGA on \name  opens up exciting opportunities~\cite{deeplearning} for exploring machine learning algorithms on-board. This would allow researchers to explore trade-offs between the power overhead of running an on-board classifier versus  sending data to the cloud. This could also enable use of high bandwidth sensors such as cameras and microphones where the power bottleneck may be communication rather than sensing. %, as well as the ability to immediately act on sensor data during network outages for example to sound a local alert or alarm.

{\bf Low power backscatter readers.} Recent work on ambient backscatter~\cite{abc,wifibackscatter,nsdi16,interscatter,fmbackscatter} aims to achieve ultra-low power communication for IoT devices. Many of these proposals require either a single-tone generator~\cite{nsdi16} or a custom receiver to decode the backscatter transmissions~\cite{3dprinted,3dprinted2,nsdi2018camera}. TinySDR can be used as a building block to achieve a battery-operated backscatter signal generation and receiver. 

{\bf Better programming interface and protocols.} In addition to IoT research opportunities, we can also improve our platform in multiple ways. TinySDR currently requires users to write Verilog or VHDL to program the FPGA and C code for programming the microcontroller. Future versions can incorporate a pipeline to use high level synthesis tools or integrate with GNUradio for easy prototyping. Further, \name  uses a simple MAC protocol for programming with a focus on using minimal system resources to allow for other custom software; however we could explore modified MAC protocols that simultaneously broadcast the updates across the network to reduce programming time.

% %-------------------------------------------------------------------------------
% \section*{Availability}
% %-------------------------------------------------------------------------------

% USENIX program committees give extra points to submissions that are
% backed by artifacts that are publicly available. If you made your code
% or data available, it's worth mentioning this fact in a dedicated
% section.

%-------------------------------------------------------------------------------

{
% \balance
\bibliographystyle{abbrv}
\bibliography{ourbib}
}
%%%%%%%%%%%%%%%%%%%%%%%%%%%%%%%%%%%%%%%%%%%%%%%%%%%%%%%%%%%%%%%%%%%%%%%%%%%%%%%%
\end{document}